# Ghostfinger: a novel platform for fully computational fingertip controllers

dr. Staas de Jong

apajong@xs4all.nl

## ABSTRACT

We present Ghostfinger, a technology for highly dynamic up/down fingertip haptics and control. The overall user experience offered by the technology can be described as that of tangibly and audibly interacting with a small hologram.

More specifically, Ghostfinger implements automatic visualization of the dynamic instantiation/parametrization of algorithmic primitives that together determine the current haptic conditions for fingertip action. Some aspects of this visualization are *visuospatial:* A floating see-through cursor provides real-time, to-scale display of the fingerpad transducer, as it is being moved by the user. Simultaneously, each haptic primitive instance is represented by a floating block shape, type-colored, variably transparent, and possibly overlapping with other such block shapes. Further aspects of visualization are *symbolic:* Each instance is also represented by a type symbol, lighting up within a grid if the instance is providing output to the user.

We discuss the system's user interface, programming interface, and potential applications. This is done from a general perspective that articulates and emphasizes the uniquely enabling role of the principle of computation in the implementation of new forms of instrumental control of musical sound. Beyond the currently presented technology, this also reflects more broadly on the role of Digital Musical Instruments (DMIs) in NIME.

## ACM Classification

• Human-centered computing~Haptic devices
• Human-centered computing~Visualization techniques
• Applied computing~Sound and music computing

## 1. INTRODUCTION: THE NEED TO ENABLE THE IMPLEMENTATION OF FULLY COMPUTATIONAL FINGERTIP CONTROLLERS

If asked to describe the system presented in this paper to someone using just one sentence, it might well be: "We present a system for running fully virtual fingertip controllers: virtual in terms of their auditory aspects, their tangible aspects, and their visual aspects." We would choose the term "virtual" here as its broadly shared meaning would quickly convey an idea of the sort of capabilities offered by the technology. However, the real motivation and potential of the technology in question is only accurately described using the word "computational". We will now first discuss why this is so, also because this reflects more broadly, beyond the currently presented system, on the role of Digital Musical Instruments (DMIs) in NIME.

### 1.1 Virtual versus computational

The existence of DMIs depends on the existence of computational transducer systems. A seminal example of this was the coupling of electronic digital computer to electric loudspeaker in the 1950s [6]. Since then, it has been made possible to induce more and more aspects of human perception that are relevant to musical control via computer-controlled transducer output. This came to include aspects of 3D vision and aspects of touch, which enabled the implementation of virtual musical instruments (VMIs) [9] [12].

Here, by its direct meaning, the word "virtual" does a good job of communicating the considerable capability of such systems to convincingly mimick aspects of reality. However, for the same reason, it is not a term that captures the full potential of the underlying transducer systems. For example, computer-generated audio output can be used to imitate acoustic pianos, as happens in digital pianos; but it can also be used to make previously unknown timbres heard, as in the case of granular synthesis [10]. Also, more fundamentally, even the very experience of hearing a stable sine wave was only made possible through the use of digital wave tables [8]. Computer-generated visual output similarly can, for example, be used to imitate the patching of analog modular synthesizers, as was the case in the Nord Modular GUI [3]; but it can also be used to visualize the setup of signal processing networks in a novel way, as was done in the reacTable system [7].

For touch output, too, a set of contrasting examples could be given (e.g. drawing from recent NIME work such as reported in [1], [2], and [11]) to illustrate the same general point: The potential of computer-driven transducer output is not only to *mimick* reality, but also to *extend* it: to implement novel forms of music making. This ongoing process of innovation is at the heart of NIME research, and it is therefore worthwhile to consider its fundamentals.

### 1.2 The principle of computation as a fundamental enabler for NIME

If we zoom out to the level of human music making in general, it turns out that, like other natural phenomena, we can often describe it as a causal chain: Human actions make changes to a sound-generating process, resulting in heard sound, which induces musical experiences within the brain (see Figure 1a). Without claiming applicability to the full spectrum of human musical activity, where it is present, each component of this causal chain is subject to empirical investigation, and potentially, intervention.

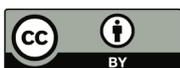

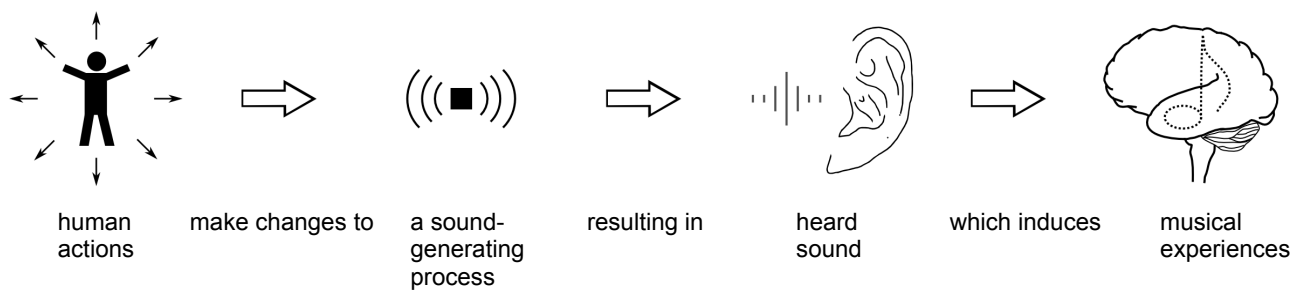

**Figure 1a** *Music making. The instrumental control of musical sound can be studied as a causal chain, wherein each component is subject to empirical investigation.*

This view on the instrumental control of musical sound then also implies a view on what it is to create a musical technique or instrument: This is to set up a causal relationship, between aspects of human action, and changes in heard musical sound (see Figure 1b).

Here, when computational technologies are given (part of) the central role of causally linking human action to heard musical sound, a unique advantage appears: Unlike earlier mechanical and electromechanical technologies, automata that are Turing-complete, when combined with transducers, inherently minimize constraints on implementable causations (see Figure 1c). This view on how the principle of computation explicitly supports the innovation of musical control was formulated and applied in [5]: There, hitherto unused aspects of human action and perception were identified and then targeted with newly developed Turing-complete transducer systems, which was then verified to yield new forms of instrumental control of musical sound.

Motivated by this research strategy, we decided to implement a novel computational transducer system targeting aspects of human action and perception that are introduced next.

### 1.3 Ghostfinger: focusing on up/down fingertip movement

Fingertip use is extremely important to the instrumental control of musical sound: The operation of many widely used musical instruments depends on it. This includes aerophones, such as flutes, oboes, clarinets, ocarinas, horns, pipe organs, trumpets, tubas, and saxophones. It also includes membranophones such as frame drums, and chordophones, such as acoustic guitars, solid-body electric guitars, violins, cellos, and (forte)pianos. More recently introduced examples are electronic instruments such as the analog synthesizer and digital sampler, as well as personal computers in their various form factors.

Moreover, in all of the examples mentioned in the previous paragraph, fingertip use contains a common component. Whether the fingertip is used to open and close holes and valves on aerophones; or to strike pads and membranes; or to tap and press on sensor surfaces; or to press strings against instrument bodies; or to perform press/release cycles on push-buttons, computer keys or the keys of piano-type keyboards: in all of these cases there is fingertip movement that can be characterized as approximating a single path of movement, at right angles with a surface, and extending across at most a few centimeters. This is illustrated in Figure 2.

For the person performing it, this up/down fingertip movement relative to a surface is associated with relatively precise control costing relatively little effort. It can also be performed at relatively high speeds, and by all fingers separately and simultaneously. This allows for making more changes to a sound-generating process over a given time period. All of these positive properties for control may help explain why specifically this type of anatomical movement has been widely used in the instrumental control of musical sound, across many cultures, and from prehistory to the present.

Because of the reasons given above, the Ghostfinger system focuses on up/down fingertip movement relative to a surface. It does so in a way aiming to preserve the fundamental advantage of computation: to explicitly minimize constraints on implementable causations between aspects of human action and changes in heard musical sound. To then characterize the forms of control that this will enable, it seems appropriate to use the term "computational fingertip controller" rather than the term "computational fingertip instrument", as in the presented prototype, a single fingertip is used for control. On the other hand, since the proposed transducer system covers hearing, touch, and vision – main areas of human action and perception via which traditional musical instruments also enable forms of control – it does not seem exaggerated to call the resulting, algorithmically implemented fingertip controllers *fully* computational.

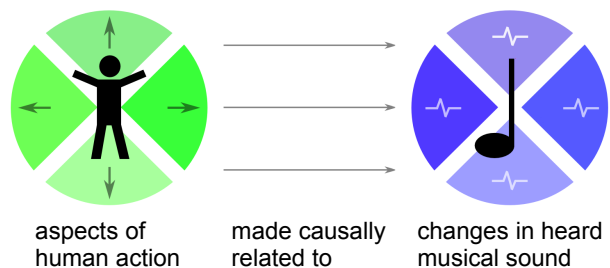

**Figure 1b** *Making music making. To create a musical technique or instrument is then to set up a causal relationship: between aspects of human action, and changes in heard musical sound.*

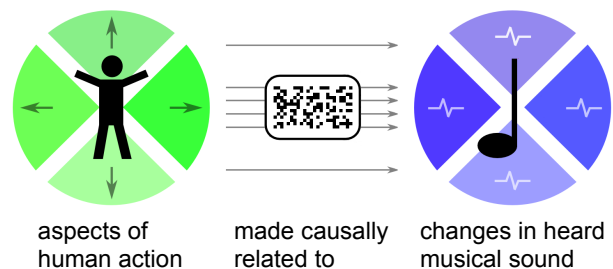

**Figure 1c** *Computational transducer systems have a unique role to play in this, because the principle of computation itself explicitly minimizes constraints on implementable causations.*

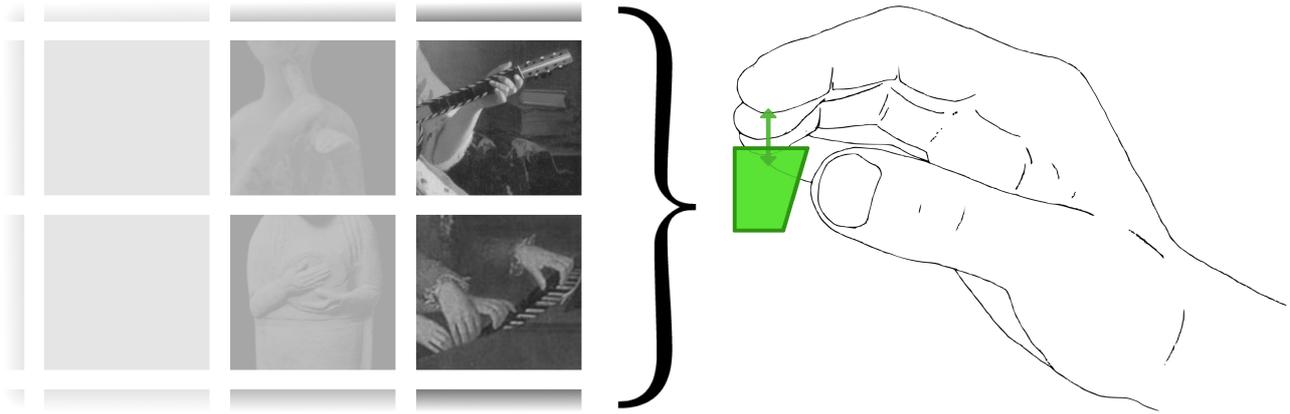

**Figure 2** *Left: examples of fingertip use on a flute, frame drum, guitar, and piano. These are shown as part of a much wider range of examples, in which musical instruments from prehistory onward have shared a common component: up/down fingertip movement relative to a surface. This is visually summarized to the right, using an intermediate hand posture, and with an example surface (as wide as the fingers) seen from below.*

## 2. THE GHOSTFINGER USER INTERFACE

The overall user experience offered by the Ghostfinger system can be described as that of tangibly and audibly interacting with a small hologram. This is implemented starting from the fingertip, where an adjustable attachment keeps a rigid transducer surface pressed against the fingerpad (see Figure 3). The hand, meanwhile, typically rests on the device surface, placing the fingertip straight above an aperture in the hardware casing. Cyclotactor technology placed beneath this aperture tracks fingertip position while projecting an attracting, rejecting, or zero force [4]. Both movement input and force output happen along a $z$ axis perpendicular to the device surface.

The fingertip transducer technology completely avoids the use of connected mechanical parts moving relative to the target anatomical site. (Detailed information about the basic hardware components of the cyclotactor subsystem can be found in [5].) This principle supports precise output to somatosensory perception, and user touch can include aspects of spatial haptic perception as well as accurate mechanical wave output across the frequency ranges involved in fingertip vibration perception. However, actually implementing such I/O requires additional higher-level algorithms, and everything presented in this paper therefore was not already covered by the cyclotactor subsystem itself.

In addition to touch stimuli, the user receives auditory and visual stimuli, also computed in real time. Audio output happens using a built-in speaker or via headphones, and visual output using a glasses-free stereoscopic 3D display. This display is located straight in front of the user's head and eyes, providing separate output to each eye via an oblique viewing angle (see Figure 3).

In this way, the user perceives, firstly, a thin grey square frame that is horizontally flush with the rest of the device surface. By its outline, this static frame visually represents the device surface area around the cyclotactor aperture, which actually is located some centimeters to the right.

Apparently floating above the static frame at surface level is a second, thicker grey square frame. This shape acts as a cursor tracking vertical fingertip movement: It visuospatially matches the current distance above surface of the fingerpad transducer. The apparent dimensions of the cursor frame precisely match those of the fingerpad transducer surface. The cursor shape differs, however, in having an open center, and this preserves an important advantage of not having touch and visual I/O spatially coincide: The absence of visual occlusion by parts of the user's hand allows for a potentially larger visual display area.

This area is used to show floating blocks, varying in their number, vertical positions, heights, colors, and levels of transparency. By their presence, the blocks indicate areas where user actions will encounter non-zero touch output. Figure 4 gives an example of this, showing visual output for a static configuration of blocks, over time, as the user performs a downward fingertip movement: At $t_1$, the fingertip is still quite some distance above the active area; at $t_2$, it has come near; at $t_3$, the fingertip enters the active area; and at $t_4$, it is inside.

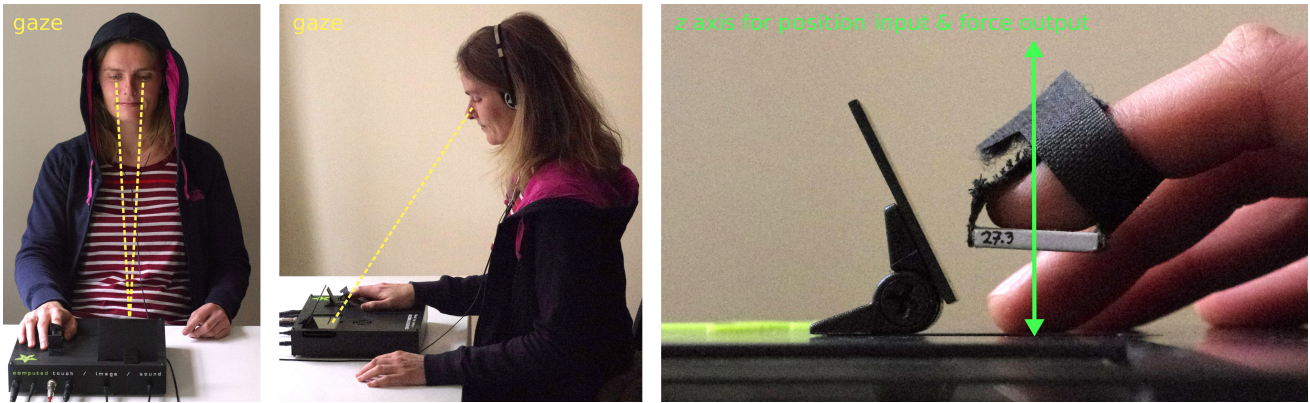

**Figure 3** *The Ghostfinger user interface.*

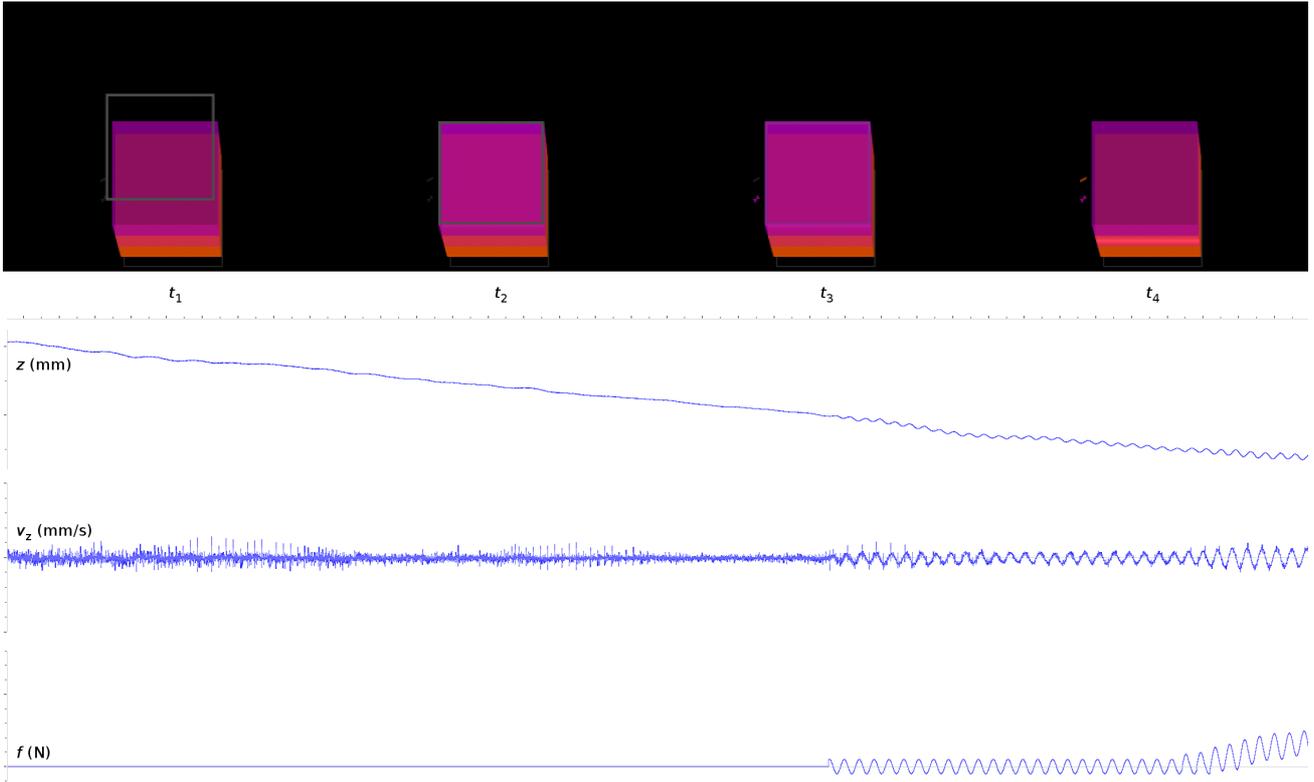

**Figure 4** *Ghostfinger example I/O. Top: right-eye visual output (to be viewed at an angle: see Figure 3). Bottom: corresponding position input, speed input, and force output, occurring along the z axis over time.*

## 3. THE GHOSTFINGER PROGRAMMING INTERFACE

In terms of its programming interface, the current implementation of the Ghostfinger system offers:
• a set of primitive types to construct algorithms for computed fingertip touch – including their automatic to-scale stereoscopic 3D visualization;
• a set of primitive types to construct forms of real-time fingertip control;
• facilities for data recording.

Table I lists the computed touch primitives, which are implemented as an abstraction layer on top of the I/O in terms of newtons and millimeters provided by the cyclotactor subsystem. Primitive types are dynamically instantiated and parametrized at runtime, by making calls in the SuperCollider language.

Each type has a limited set of runtime parameters, always including $z$ base position and size, so as to set up the range where the primitive instance may compute non-zero force output. For the "monoforce" type, this output simply is a single force level, uniformly present across the given range. For the "linear ramp" type, force output is computed from a linear gradient over distance between two arbitrarily chosen points in the overall force output range. The "dashpot" type implements a requested level of viscosity across its spatial range. A "directional dashpot" does the same, but for viscosity that is present in one movement direction only. Finally, a "force wave" instance computes a bipolar force sine wave over time, of precisely settable frequency and amplitude.

Under the hood, the subsystem automatically visualizing computed touch primitives is implemented using OpenGL, and fully custom from the pixel shaders on up. This also includes atypical geometric 3D projection computations, duplicated for each eye, and necessary because instead of like the usual window, the electronic screen hardware faces the user like an obliquely viewed platform (see Figure 3).

| symbol | name | runtime parameter (*unit*) |
|---|---|---|
|  | (*shared by all*) | z base position (*mm*)<br>z size (*mm*) |
| – | monoforce | z force (*N*) |
| ╱ | linear ramp | z force at base (*N*)<br>z force range (*N*) |
| ╪ | dashpot | z viscous damping (*N*/(*mm*/*s*)) |
| ╧ | directional dashpot | z viscous damping (*N*/(*mm*/*s*))<br>z direction (*bit*) |
| ✦ | force wave | frequency (*Hz*)<br>amplitude scaling (*N*) |

**Table I** *Primitive types for computed fingertip touch.*

Fundamentally, the visualization subsystem listens in on ongoing instantiation, parametrization and termination events, and visualizes the *types* and *runtime parameters* of the computed touch primitives that currently exist. Each primitive instance is represented by a block shape, placed along the vertical path of the cursor. The base and top surfaces of a block have the same perceived horizontal dimensions as the cursor, and vertically match with the current active range for touch output. Block color displays primitive type. The state of the remaining, type-specific runtime parameters (see Table I) is displayed by block transparency. Here, as a general rule, full transparency always means 0 N output, and increasing opacity indicates an increasing "strength".

To aid the spatial perception of block shapes and their sizes, the presence of static directional lighting is simulated. In addition to this, the cursor is simulated to light up its immediate surroundings. This to enhance cursor visibility inside opaque volumes, and to generally highlight the (imminent) passing of active range boundaries (see Figure 4).

Primitive instances each add to the overall force computation, and are independently parametrized to do so. Therefore, it may well be that two or more primitives come to spatially overlap. The visualization subsystem detects this, and renders corresponding overlapping block segments to reflect it. The RGB color of such a block segment is computed as an average of the type colors of the overlapping primitive instances, with each type color weighted by its instances' current opacities. So, if two "strong" primitive instances combine, their overlap block segment will show an intermediate color; but if a "weak" instance overlaps a "strong" one, the color of the former may only slightly tinge that of the latter. In general, for $n > 0$ overlapping primitives of which at least one has a non-zero opacity (or "A value"), the combined $(r, g, b, a)$ tuple is given by

$$\sum_{i=1}^{n} \left( a_i \;/\; \sum_{j=1}^{n} a_j \times (r_i, g_i, b_i) \right) \frown \max_{k=1}^{n} a_k .$$

The opacity of an overlap block segment is therefore computed as the maximum of the overlapping primitives' opacities. This ensures that if a visible primitive instance overlaps with others that are at maximum transparency – and which therefore, are contributing 0 N output – the color and opacity looked at will be the same as for the primitive instance on its own. Figure 4 shows a basic example of two primitive instances of different type, spatially overlapping, which yields a third, middle block segment.

The spatial display via cursor and blocks is complemented by symbolic display, visible as a sign grid extending outward from the static floor frame toward the left and right. Here, each primitive instance is represented by a corresponding type symbol (see Table I) that lights up if the fingerpad transducer is within the instance's active range. Figure 4 shows an example of this for a force wave and a linear ramp instance. During the programming and trying out of algorithms for dynamic haptics, the sign grid supports unambiguous verification of which primitive instances are being activated during fingertip actions – especially as the instances' spatial dimensions and opacities decrease, and their numbers increase. The Ghostfinger system currently supports having up to 160 concurrently running primitive instances.

When starting algorithmic construction from a single primitive instance, fingertip movement staying within the active range may bring about aspects of action and perception that are experienced as distinct from those induced while traversing the lower or upper range boundaries. For example, when entering a dashpot or upward-only directional dashpot instance from above, the sudden activation of its viscosity may yield an impact sensation.

In addition to the set of primitive types for computed touch, the Ghostfinger system also offers a set of primitive types for implementing forms of real-time control via the fingertip.

These types of Degree-Of-Freedom (DOF), too, are dynamically instantiated and parametrized at runtime. The currently implemented selection is listed in Table II. Here, some types provide updates continuously, tracking fingerpad transducer position, average position, average absolute deviation from average position, or speed. Other types signal discrete events, such as fingertip entry into and exit from a given spatial range, or the fingertip's speed as it passes a positional threshold in a given direction.

| DOF name (*unit*) | runtime parameter (*unit*) |
| --- | --- |
| inside (*bit*) | z base position (*mm*) <br> z size (*mm*) |
| z relative position (*mm*) | z base position (*mm*) |
| z avg. relative position (*mm*) | z base position (*mm*) <br> averaging period (*ms*) |
| z avg. abs. dev. from avg. (*mm*) | averaging period (*ms*) |
| z speed (*mm/s*) | – |
| downward pass (*mm/s*) | z inclusive threshold (*mm*) |
| upward pass (*mm/s*) | z inclusive threshold (*mm*) |

**Table II** *Primitive types for implementing fingertip control.*

Designing fingertip control actions using Ghostfinger will often involve spatially lining up DOF primitive instances with ones for computed touch. For example, the threshold location for which a "downward pass" DOF instance will trigger and report movement may be placed at the top boundary of an upward-only directional dashpot instance: Then, once a tangible impact sensation unfolds there, the speed of the fingertip (in millimeters per second) can also control e.g. the loudness (in decibels) of some heard sound, also computed in real time.

Finally, the Ghostfinger system also offers facilities for recording I/O data over time. This is done at the temporal resolution of the cyclotactor subsystem, with amplitude values always having physical units. Figure 4 shows an example of this, with recorded position, speed, and force data reflecting a downward fingertip movement into the active ranges of first, a force wave primitive, and then also a linear ramp primitive that is implementing a spring.

## 4. DISCUSSION: POTENTIAL APPLICATIONS

We will now mention some potential applications of the Ghostfinger platform, listed along an axis of user experience that goes from expert to novice. This simultaneously means going reciprocally along an axis of "amount of work needed": the less expert the user, the more work that still needs to be done to make the proposed application a reality.

- *As a platform for basic research:* Ghostfinger can be used to create and research novel forms of instrumental control of musical sound involving the fingertip. This research can also include controlled experiments quantifying the effects of different haptic conditions on musical control outcomes.

- *As a platform for constructing controllers:* A future, more embedded version of Ghostfinger might be used by researchers and advanced students to freely construct fingertip controllers that employ highly dynamic haptics, without such users having to deal with programming direct transducer I/O. The video in the Appendix shows examples that relate to this use case.

- *As a platform for using controllers:* After even more development, a variant of Ghostfinger technology might offer end users sets of

ready-made musical controllers, without requiring any programming on their side. Out of the box, such controllers should communicate using standard musical I/O protocols such as OSC and MIDI, so as to easily fit in existing contexts for digital music making. Added value might lie in adding a range of simulated fingertip actions that could be switched between on a whim; and also in enabling completely new ways of fingertip control.

## 5. CONCLUSION

We will conclude this paper by reviewing some important aspects of the functionality of the Ghostfinger user interface and programming interface.

The automatic visualization of the floating cursor, via its perceived open-frame shape and spatial movement, immediately, non-verbally, and continuously communicates to the user what type of user actions are being picked up: vertical fingertip movements. This makes the user interface much more self-explanatory than in the absence of such a cursor. Equally importantly, the cursor also immediately communicates what types of movement are *not* being used: movements such as fingertip pitch, roll, and yaw. In this way, spatiovisual display prevents user confusion which, earlier, did occur during informal testing with musicians of just the cyclotactor system. (This confusion, on reflection, seemed to prohibit open-ended yet still intuitive use of the technology, which directly motivated developing the additional visual output presented in this paper.)

The programming primitives and underlying transducer I/O of the Ghostfinger system enable implementing widely varying and highly dynamic haptic conditions for up/down fingertip movement. In the user, this can induce aspects of passive touch and active touch, including exterospecific components yielding spatial haptic perception. Moreover, higher-level algorithms can then change these induced exterospecific components in response to specific human motor activity, thereby effectively conveying the occurrence of not just active touch, but tangible manipulation.

By adding or re-using DOF primitive instances, such algorithms of active touch and tangible manipulation can then be extended to also compute sound in response to user actions, thereby implementing tangibly distinct modes of musical control. In turn, algorithms at an even higher level may then be used to freely and rapidly switch between different modes of control, and potentially do so many times within the timespan of a single performance.

This means that for the user, the set of possible control actions may become highly dynamic over time. Given that, so far, we have mostly reviewed the potential of computational touch and audio I/O, this raises a question: How does the user know what control actions currently are possible?

The automatic visualization built into the Ghostfinger system addresses this question without relying on visually mimicking pre-existing objects or forms of manipulation. Instead, transparently colored and spatially overlapping floating blocks visualize the runtime instantiation and parametrization of haptic primitive instances that together combine to form the specific current context for control actions. This open-ended form of spatial visual display is complemented by the simultaneous symbolic display of active primitive instances visible in the sign grid.

Finally, we turn to the programming interface specifically. The goal was to make it powerful, yet easy to use: The set of primitive types for computed touch and DOF extraction is small, and each type is conceptually simple and has only a few runtime parameters. Simultaneously, each type can be instantiated many times over, and can be freely combined with any other type. Overall, the type set was designed to be expressive, in the sense that resulting, more complex algorithms may implement many different forms of touch and control.

Given our reflection in Section 1 on the general role of computation in the implementation of forms of instrumental control, an important issue here is preserving the full computational potential corresponding to minimized constraints on the potential implementation of causal relationships between aspects of human action and changes in musical sound. As a prototype means for preserving access to the full range of potential algorithmic implementation, "*z* relative position", "*z* speed", and "monoforce" primitives can be used as a wrapper for the underlying transducer I/O.

## APPENDIX

A 7½-minute video introducing and demonstrating the Ghostfinger prototype is at https://youtu.be/ahw9630FLgU.